 \documentclass[aps,prb,twocolumn,groupedaddress,showpacs]{revtex4}
\usepackage{graphicx}       
\usepackage{dcolumn}        
\usepackage{bm}             

\begin{document}

\title
    {
 Photovoltage Dynamics of the Hydroxylated Si(111) Surface Investigated by Ultrafast Electron Diffraction
    }

\author{Ryan A. Murdick}
\affiliation{Department of Physics and Astronomy,
             Michigan State University,
             East Lansing, Michigan 48824-2320 }

\author{Ramani K. Raman}
\affiliation{Department of Physics and Astronomy,
             Michigan State University,
             East Lansing, Michigan 48824-2320 }

\author{Yoshie Murooka}
\affiliation{Department of Physics and Astronomy,
             Michigan State University,
             East Lansing, Michigan 48824-2320 }

\author{Chong-Yu Ruan}
\email[]{Email: ruan@pa.msu.edu}
\affiliation{Department of Physics and Astronomy,
             Michigan State University,
             East Lansing, Michigan 48824-2320 }



\begin{abstract}
We present a novel method to measure transient photovoltage at nanointerfaces using ultrafast electron diffraction. In particular, we report our results on the photoinduced electronic excitations and their ensuing relaxations in a hydroxyl-terminated silicon surface, a standard substrate for fabricating molecular electronics interfaces. The transient surface voltage is determined by observing Coulomb refraction changes induced by the modified space-charge barrier within a selectively probed volume by femtosecond electron pulses. The results are in agreement with ultrafast photoemission studies of surface state charging, suggesting a charge relaxation mechanism closely coupled to the carrier dynamics near the surface that can be described by a drift-diffusion model. This study demonstrates a newly implemented ultrafast diffraction method for investigating interfacial processes, with both charge and structure resolution.
\end{abstract}

\pacs{
82.53.Mj, 
61.05.J-, 
68.49.Jk, 
82.65.+r 
}



\maketitle



The problem of directly converting solar energy into electrical energy through photocarrier generation has recently gained tremendous attention
due to new types of photovoltaic (PV) cells utilizing nanoparticles and molecular interfaces. \cite{Alivisatos,Gratzel2001,Nozik,Lewis2007,Aydil2007}
An important parameter determining the efficiency of the solar cell is the survival rate of photogenerated carriers reaching the electrodes,
which is strongly affected by surface recombination and the transport characteristics at the contacts. In a Schottky-type PV junction,
the carrier dynamics are driven by a space-charge barrier layer. \cite{Schottky} A photovoltage is generated as the photoinduced electron-hole
pairs near the semiconductor interface move to reduce the band-bending. More generally, diffusion of non-equilibrium carriers, \cite{Davydov,Mott,Tauc}
charging of the surface states, \cite{Bokor1989} and photoionization \cite{Demuth1986} are different contributors for \textit{emfs}
at semiconductor junctions, attributing to the overall magnitude and sign of the generated photovoltage. For a molecular interface,
the existence of discrete surface and/or charge transfer states are crucial in directing the carrier separation in new generation PV cells,
such as dye-sensitized TiO$_2$ mesoporous films \cite{Gratzel2001} and quantum dot solar cells. \cite{Nozik,Aydil2007} In these cases, the filling of a
surface state can strongly affect the charge distribution in the space-charge layer, modifying the transport characteristics,
as revealed in photoemission studies. \cite{Bokor1989,Kabler,Marsi} Surface charging is also essential for explaining surface
photochemical processes, including catalysis \cite{Landman} and molecular transport. \cite{Datta} Here, we use a novel diffractive
potentiometry approach to determine the transient photovoltage based on the ultrafast electron diffraction (UED) technique,
whose ability to resolve surface structure evolution has been demonstrated previously. \cite{Zewail2006,UEC_NL}

We study the photoinduced surface potential changes on a Si(111) surface terminated with hydrophilic hydroxyl (OH) groups,
a prototypical system for fabricating molecular electronic devices. \cite{Shipway, Aswal} The experiment is performed in
an optical pump\verb - diffraction probe arrangement, in which the femtosecond laser pulse is used to initiate carrier
generation near the surface and in turn, the formation of a transient surface voltage (TSV), while a charge-sensitive
electron pulse probes the ensuing carrier relaxation dynamics. A Coulomb refraction formalism is deduced to treat
ultrafast diffraction data to determine the TSV. Using a laser fluence ($F$) of 22 mJ/cm$^2$, we found a band-flattening
TSV of ~300 mV generated ~30 ps after the laser excitation.  At higher fluences ($F > 72$ mJ/cm$^2$), the TSV continued
to rise, even beyond the bandgap energy. This surprisingly large TSV is attributed to a modified barrier due to a
non-equilibrium surface charge migration, induced by photoexcitation. \cite{Marsi} We observed picosecond charge injection
and relaxation dynamics with timescales similar to results obtained using ultrafast photoemission studies of surface carrier
dynamics in a vacuum cleaved Si(111)-($2 \times 1$) surface. \cite{Bokor1989} This rapid recovery can be understood based on
a drift-diffusion model that couples surface carrier dynamics with transport in the space-charge layer. This suggests that
the hydroxylated silicon surface posseses a lower concentration of trap states, ideal for mediating molecular electronic transport study. \cite{Aswal}

The hydroxylated Si(111) surface was prepared {\em ex situ} using a wet chemistry method. The as received Si(111) wafer
(Silicon Quest Intl., p-type, roughness $<$ $12$~$\mu$m) was treated with a modified RCA procedure. \cite{Puotinen}
The native oxide layer, which usually contains a high density of trap states, was removed and an ultrathin ($<$ 5 nm)
thermal oxidation layer with high electric quality was grown. First, the wafer was immersed in H$_2$SO$_4$/H$_2$O$_2$ (7:3)
solution at $90\,^{\circ}{\rm C}$ for 10 min to remove surface contaminants. The surface was then etched with a saturated
NH$_4$F solution \cite{Becker} to remove the native oxide layer, followed by a bath in NH$_4$OH/H$_2$O$_2$/H$_2$O (1:1:5)
at $80\,^{\circ}{\rm C}$ for 20 min to further remove inorganic residues. Finally, a thin layer of oxide that affords a
hydroxylated surface was grown by immersing the wafer in HCl/H$_2$O$_2$/H$_2$O (1:1:6) at $80\,^{\circ}{\rm C}$ for 10 min.
The wafer was rinsed with deionized water ($17.8$ M$\Omega\cdot$cm) for 10 min at the end of each cycle. This procedure
yields a density of $\approx 10^{15}$ cm$^{-2}$ silanol groups (Si-OH) on the surface. \cite{Aswal}

\begin{figure}
\includegraphics[width=1.0\columnwidth]{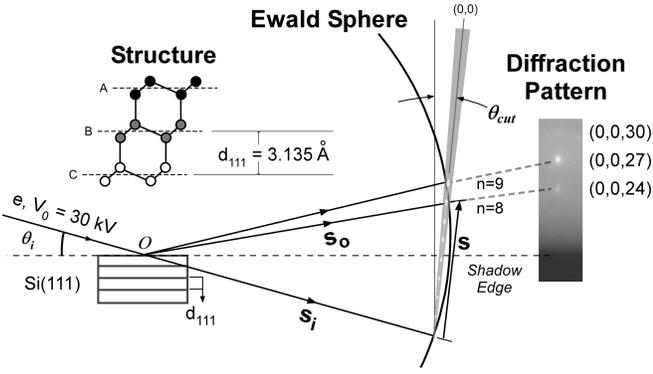}
\caption{Electron scattering geometry. The incident beam, \textbf{s}$_i$ (energy $eV_0$, incidence angle $\theta_i$,
Ewald sphere origin \textit{O}) scatters off the Si(111). Constructive interference occurs when the scattered beam (\textbf{s}$_o$)
crosses an intersection of the Ewald sphere with a reciprocal lattice rod. Here, surface steps from the Si(111) cleave,
tilt the reciprocal lattice rods by $\theta_{cut}$, resulting in interference maxima below the projected Zeroth Order Laue Zone.
From observable peaks (0,0,24), (0,0,27), and (0,0,30), we measured $\Delta s = 1.985$ \AA,~implying interlayer spacing
$d_{111} = 2 \pi/{\Delta s} = 3.165$ \AA~and $\theta_{cut} \approx 1.35\,^{\circ}$. The ABC stacking of Si(111) diamond lattice
has lattice constant 9.406 \AA, implying an interlayer spacing of $d_{111} = 3.135$ \AA, in agreement with our measurements.\label{Fig1}}
\end{figure}

The experiment was conducted in the ultrafast electron crystallography setup, \cite{UEC_NL} employing the reflection geometry
to gain surface sensitivity. As described in Fig.~\ref{Fig1}, the incident probe electron has a kinetic energy $K_0$ = $eV_0$,
where $V_0$ = 30 kV is the accelerating voltage, corresponding to an incident reciprocal wavevector $s_i$ = 90 \AA$^{-1}$.
At an incidence angle, $\theta_i$ = $6.8\,^{\circ}{\rm}$, the Ewald sphere intersects the (0,0) reciprocal lattice rod of the
Si(111) substrate near momentum transfer wavevector $s \approx 18$ \AA$^{-1}$, producing an in-phase (0,0,27) diffraction
maximum on the screen. Since the Si(111) substrate has a slight miscut, ($\theta_{cut}$ $\approx$ $1.35\,^{\circ}$),
the diffraction maxima appear slightly below the projected Zeroth Order Laue Zone. Weak (0,0,24) and (0,0,30)\cite{PEAK_SN}
diffraction maxima arising from the surface roughness were also observed.  These maxima are associated with the interference pattern
of Si(111) planes, evident from $\Delta s$ = 1.985 \AA$^{-1}$ between neighboring peaks satisfying the reciprocal relationship
$d_{111}$ = 2$\pi/\Delta s$ with an accepted $d_{111}$ value of 3.135 \AA. The electron penetration depth in silicon is estimated
to be $\approx$ 2 nm based on the Scherrer criterion, \cite{Patterson} using the width of the (0,0,27) diffraction maximum.

\begin{figure}
\includegraphics[width=0.7\columnwidth]{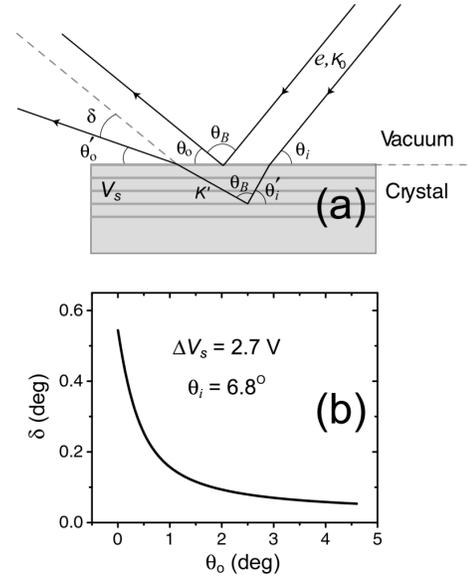}
\caption{ (a) Ray traces of the electron path with and without a surface potential ($V_s$) present. $V_s$ induces a shift $\delta$ in the diffraction maxima. (b) The shift in diffraction peak maxima as a function of $\theta_o$. Lower order diffraction maxima shift more than higher orders in the presence of a surface potential. \label{Fig2}}
\end{figure}

To determine the surface potential, we examine the Coulomb refraction introduced by a surface potential for electron diffraction,
as shown in Fig.~\ref{Fig2}(a). The surface potential, $V_s$, alters the kinetic energy of the electrons submerged in the crystal,
$K'$ = $e(V_0+V_s)$, without changing the transverse momentum, thereby causing a refraction at the interface.
The index of refraction can be described by $n_e = \sqrt{(V_s + V_0)/V_0}~$. \cite{Wang1996} For surface diffraction,
this alters the electron incidence angle towards the lattice plane, thus introducing a net angular shift ($\delta$) of
the diffraction maximum. We calculate the shift arising from a surface potential to be

\begin{equation}
\frac{V_s}{V_0} = \frac{(\theta_o \delta - \theta_i \theta_o + \delta^2/2)^2 - \theta_i^2 (\theta_o + \delta)^2 }{(\theta_i + \theta_o)^2},
\label{Eqn1}
\end{equation}

\noindent where all symbols are defined in Fig.~\ref{Fig2}(a). In the optical pump\verb - diffraction probe experiment, $V_s$ is replaced by the transient surface voltage, $\Delta V_s$, to evaluate the photoinduced refraction shift of diffraction maxima at $\theta_o$. The dependence of $\delta$ on $\theta_o$ is shown in Fig.~\ref{Fig2}(b), calculated using Eqn.(\ref{Eqn1}) with our incidence angle ($\theta_i$ = $6.8\,^{\circ}{\rm}$) and $\Delta V_s = 2.7$ volts. This refraction shift has a `non-structural' characteristic, exhibiting a more significant deviation at smaller values of $\theta_o$. This differs fundamentally from structure-induced shifts, which behave oppositely. For small angle diffraction the Bragg law yields a monotonic increase in $\delta$ as a function of $\theta_o$ for a given structural change $\Delta d$, described by $\delta/(\theta_o + \theta_i) = \Delta d/d$, i.e. the higher order diffraction maxima experience proportionally larger shifts than the lower order ones. This clear distinction allows us to separate Coulomb-induced effects from structural changes by analyzing multiple diffraction maxima.
To demonstrate the principle, we will follow both (0,0,24) and (0,0,27) diffraction maxima to examine the TSV dynamics.

\begin{figure}
\includegraphics[width=1.0\columnwidth]{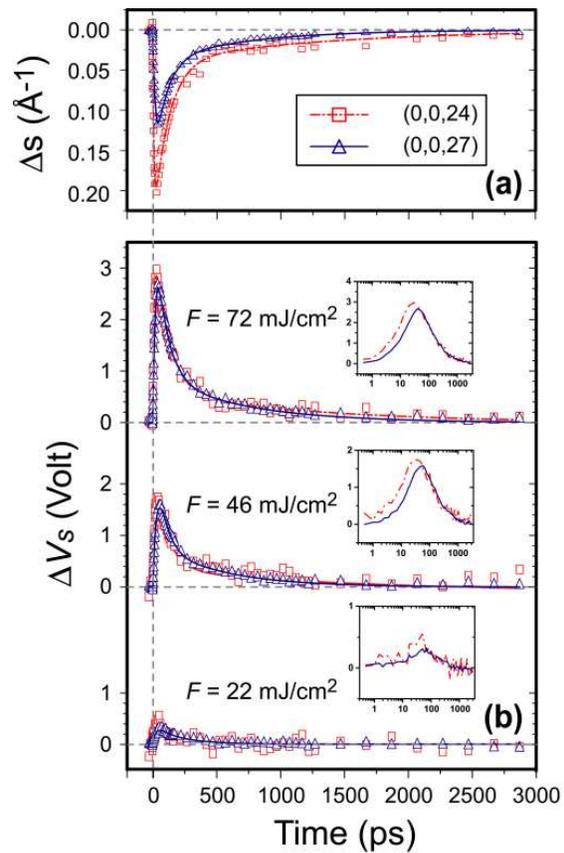}
\caption{(Color online) (a) Transient shift in diffraction maxima induced by the TSV. The shift in peak (0,0,24) nearly doubles that of (0,0,27), indicating the presence of a TSV. (b) The TSV for the laser fluences employed. Insets: semi-log TSV, demonstrating the contrasting behavior in the TSV rise for the given peaks.\label{Fig3}}
\end{figure}

The characteristic features associated with the Coulomb effect were observed experimentally as shown in Fig.~\ref{Fig3}(a). Using a laser fluence of 72 mJ/cm$^2$, the time-dependent shifts of diffraction maxima (0,0,24) and (0,0,27) were followed by adjusting the arrival time of the electron probe pulse relative to that of the exciting laser pulse. The shift of the (0,0,24) diffraction maximum in reciprocal space is markedly greater than that associated with (0,0,27), pointing to a Coulomb effect. We expect a negligible structure-induced shift because of the small absorption coefficient in silicon ($\alpha^{-1} \approx 1 \mu$m) for the near infrared excitation ($h \nu = $1.55 eV), indicating that the thermal energy deposition is spread out over a large volume. The maximum temperature rise in Si can be estimated from $\Delta T_L = C_L^{-1} \alpha F (1-R)(h \nu - E_g)/h \nu + C_L^{-1} \Delta n_{e-h} E_g$, where $R = 0.37$ is the reflectivity,
$C_L=2.08 \times 10^{6}$ Jm$^{-3}$K$^{-1}$ is the lattice heat capacity, and $\Delta n_{e-h}$ is the carrier density drop due to Auger recombination. \cite{Chen} The laser fluences we applied ranged from 22 to 72 mJ/cm$^2$, which give rise to only moderate increases in lattice temperature, about 20 to 170 K, respectively, above the initial temperature, $T_0$ = 300 K. Hence the changes induced by lattice heating, $\Delta s/s$, are on the order of $10^{-4}$ based on the thermal expansion coefficient for Si. \cite{Glazov} In contrast, the carrier generation is more significant, ranging from $5 \times 10^{20}$ to $ 1.4 \times 10^{21}$ cm$^{-3}$, which corresponds to an increase of several orders of magnitude from the intrinsic carrier concentration ($10^{15}-10^{16}$ cm$^{-3}$) within the first ps. \cite{Chen} It is expected that such a large increase in carrier density will easily flatten out the initial surface band-bending, which is $\approx 0.3$ V, \cite{Widdra} leading to a change $\Delta s/s \approx 7.8 \times 10^{-3}$ for the (0,0,27) maximum. Indeed, a band-flattening TSV of 0.38 V was determined at the lowest fluence applied ($F$ = 22 mJ/cm$^2$). However, TSVs higher than the initial band-bending were also observed; 1.74 and 2.97 V for fluences of 46 and 72 mJ/cm$^2$, respectively. In Fig.~\ref{Fig3}(b), the TSVs deduced from Eqn.(\ref{Eqn1}) are consistent for both the (0,0,24) and (0,0,27) maxima, for all three fluences, demonstrating the robustness of the diffractive TSV measurment.

The significant photovoltage induced by the high fluence femtosecond laser cannot be explained merely by a barrier-layer effect, where photoinduced e-h pairs near the semiconductor interface separate to screen the surface charges responsible for the band-bending. Without modifying surface charge population, this theory would predict a maximum TSV corresponding to the dark current band-bending \cite{Kronik} and the decay time of the TSV would be comparable to the characteristic bulk recombination time ($\approx 10-100$ ns for Si). This is not what is observed here. In contrast, we have witnessed a much more rapid process, with the TSV surging within 30 ps following laser excitation and a decay on the order of 100 ps for each of the three different fluences, without any saturation, as shown in Fig.~\ref{Fig3}(b). Enhanced surface band-bending was also observed by Marsi \textit{et al}.~in studying the transient charge distribution at the SiO$_2$/Si interface following UV free electron laser excitation using time-resolved pump-probe core photoemission spectroscopy. \cite{Marsi} The cause of the enhancement is attributed to electron diffusion into the thermally grown SiO$_2$ overlayer and subsequent accumulation at its surface. The recombination of the excess electrons was found to be slower with thicker oxide layers. For 12 \AA ~oxide thickness, the surface charge recombination was found to be comparable to the typical excess carrier recombination time in the Si(111) space-charge layer, of the order of 100 ns. For very thin (a few \AA) oxide layers, a very effective `surface' recombination process was observed within their time resolution of $\approx$ 200 ps. These observations are consistent with a picture in which the surface recombination rate is determined by the separation between the excess surface electrons and excess holes in the Si space-charge layer. Indeed, much shorter recombination time is found for the decay of the surface state population in a vacuum-cleaved Si surface. \cite{Bokor1989} This speed-up in surface recombination for an ultrathin oxide layer ($< 1$ nm) observed by Marsi \textit{et al}.~is attributed to the enhanced overlap of bulk evanescent states of Si with the surface states. In that case and in the studies reported here, the electrons accumulated at the thin oxide surface are so close to the Si space-charge layer that their role in the recombination process is similar to the normal surface state.

\begin{figure}
\includegraphics[width=0.85\columnwidth]{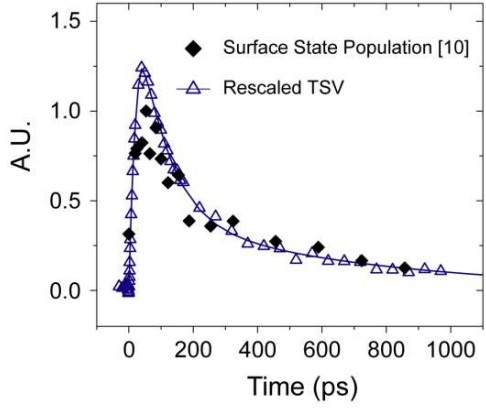}
\caption{(Color online) Transient surface state population measured by photoemission \cite{Bokor1989} compared with the measured TSV (rescaled; $F=72$ mJ/cm$^2$), demonstrating very similar decay rates.\label{Fig4}}
\end{figure}

\begin{figure}
\includegraphics[width=1.0\columnwidth]{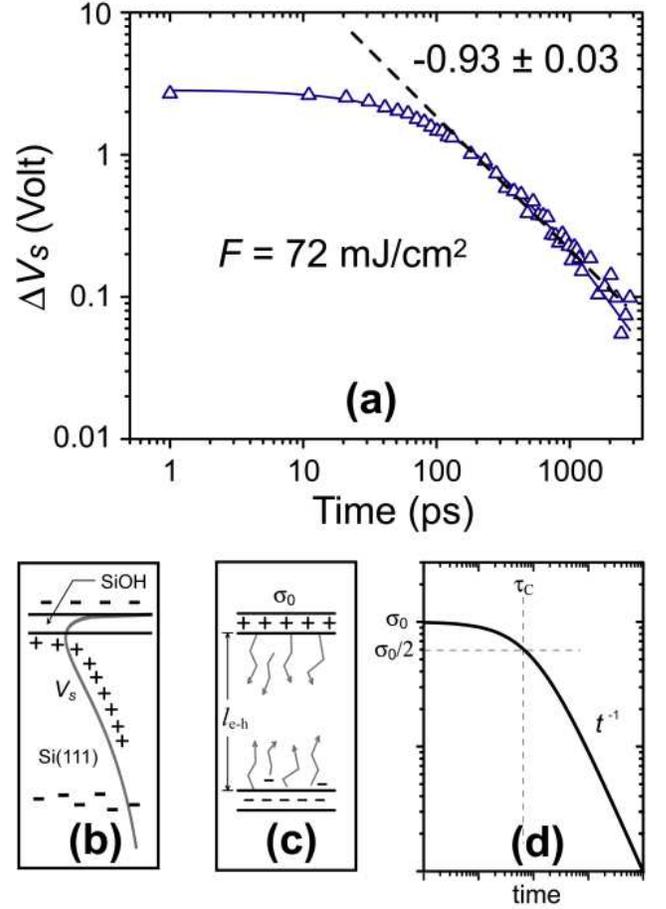}
\caption{(Color online) (a) Log-log plot of the TSV. (b) Surface potential ($V_s$) at the SiOH/Si(111) interface, depicted by solid gray line. (c) Two-slab model employed (schematic) to describe the space-charge region. (d) Carrier drift causes the surface charge ($\sigma_0$) to diminish with characteristic time $\tau_c$, after which, $\sigma_0$ falls as $t^{-1}$, obtained from two-slab model.\label{Fig5}}
\end{figure}

The decay behavior of the TSV on the hydoxylated Si(111) surface was compared with the surface state dynamics on a vacuum-cleaved Si(111)-($2 \times 1$) surface, revealed by a photoemission study (Fig.~\ref{Fig4}). \cite{Bokor1989} In this figure, the TSV is rescaled to match with the data extracted from Ref. 10. We find a striking similarity between the two, suggesting that the surface state population is in quasi-equilibrium with the bulk carrier dynamics in the space-charge layer, so long as there is strong coupling between the surface state and the bulk evanescent state. Further evidence of the strong role played by carrier dynamics on the surface recombination is revealed by the power-law decay of the TSV, shown in Fig.~\ref{Fig5}(a), where an exponent near -1 (-0.93$\pm$0.03) is ascribed to the characteristic decay of surface charges, with a time constant of 100 ps.

To describe the population decay of the surface state driven by a space-charge drift recombination, we introduce a simple two-slab model, illustrated in Fig.~\ref{Fig5}(b)-(d). The electrons accumulated at the surface lead to a rapid rise in surface potential within the thin oxide layer that decays slowly into the Si space-charge region, as depicted in Fig.~\ref{Fig5}(b). The non-equilibrium space-charge layers, with initial separated charge density $\sigma_0$, are modeled as two separate slabs, separated by a distance $l_{e-h}$ in a dielectric medium ($\epsilon \approx 13$ for Si), shown in Fig.~\ref{Fig5}(c). The power-law characteristics are manifested in the simple rate equation $d \sigma(t)/dt = \sigma(t)/\tau_r$, with the space-charge recombination time $\tau_r=l_{e-h}/\mu E$ depending on the transient field $E$, which is directly related to the surface charge $\sigma(t)$. Its solution $\sigma(t) = \sigma_0/(t/\tau_c+1)$ has a characteristic timescale $\tau_c = l_{e-h} \epsilon/\mu \sigma_0$, which is the time for the TSV to drop by 50\% from its initial value, and corresponds to the induction period in the log-log plot of the solution, as described in Fig.~\ref{Fig5}(d). For $t >> \tau_c$, the $t^{-1}$ behavior emerges, similar to what is observed experimentally [Fig.~\ref{Fig5}(a)]. To compare with the experimental results, we assume a linear decay of the surface potential barrier ($V_s$) over the space-charge layer $l_{e-h}$, and determine the size of the induced space-charge regime to be $l_{e-h} \approx 400$ nm (using carrier mobility, $\mu$=100 cm$^2 $V$^{-1} s^{-1}$ according to Ref. 25, our measured $\tau_c$ of 100 ps, and $V_s$ of 2 V), which is in agreement with the laser penetration depth in Si.

Other mechanisms that would also affect the TSV decay include the ambipolar diffusion of the bulk excess carriers away from the surface and the decay of the surface state charges into trap states. Here, we can exclude ambipolar diffusion as the main cause for TSV decay for two reasons. First, the characteristic diffusion time, $\tau_{diff}$, is significantly longer than what we observed ($\tau_{diff} = l_{e-h}^2/D_{e-h} \approx 700$ ps, based on the excess e-h region generated by the laser, $l_{e-h} \approx$ 1 $\mu$m, and ambipolar diffusivity $D_{e-h} \approx$ 15 cm$^2$s$^{-1}$). \cite{Chen} Second, a $t^{-1/2}$ decay is expected in the case of one-dimensional diffusion, \cite{Zewail2007_2} which is appropriate for photoexcitation when the laser spot size (600 $\mu$m) is large compared to the penetration depth ($\approx 1~\mu$m). The rapid decay of the TSV after 100 ps precludes any long-lived trap state from playing a significant role. Halas and Bokor invoked coupled drift-diffusion-Poisson equations for bulk carriers to self-consistently treat the surface state dynamics. \cite{Bokor1989} Such a calculation is able to describe the general behavior of the surface state depopulation and TSV decay (to be reported elsewhere \cite{UEC_MICKENS}). However, the essential power-law behavior can be elucidated with a simple space-charge recombination model, as described above.

While the simple Coulomb refraction formalism described in Eqn.(\ref{Eqn1}) seems to be adequate for deducing consistent TSVs from diffraction maxima, especially for the recombination dynamics, some departure between the TSVs deduced for (0,0,24) and (0,0,27) is evident at short times, as depicted in the insets of Fig.~\ref{Fig3}(b). This discrepancy is believed to be caused by a non-equilibrium barrier layer, which deviates significantly from a homogeneous potential model assumed in Eqn.(\ref{Eqn1}) within the electron probed volume. More rigorous treatment of the Coulomb refraction formalism with a non-homogeneous potential is needed to fully account for the short-time dynamics.

In conclusion, we have demonstrated a new method of directly measuring transient surface potential using ultrafast electron diffraction. In comparison with other commonly used surface imaging techniques, such as Kelvin Force and Scanning Tunneling Microscopy, we are able to examine surface potential changes from the femtosecond to nanosecond timescale. This technique is complementary to ultrafast photoemission methods in which the energetics of the photoemitted electrons, rather than the potential, are directly measured. Since our employed electron probe energy is reasonably high, it allows surface potential characterization at high laser excitation fluences, which are usually inaccessible by the low energy photoemission techniques due to the space-charge effect. \cite{Gilton1990} We have revealed a regime of strong photoinduced surface voltage rise, which can serve as an impulsive current source for investigating time-dependent transport at nanointerfaces.

This work was supported by Department of Energy under grant
DE-FG02-06ER46309.


\begin{thebibliography}{31}
\bibitem[{PEA()}]{PEAK_SN}
\bibinfo{journal}{The signal-to-noise ratio for this peak was too weak to
  follow dynamically.}
\end{thebibliography}

\end{document}